**Atomic layer graphene as saturable absorber for ultrafast pulsed lasers ****

By *Qiaoliang Bao, Han Zhang, Yu Wang, Zhenhua Ni, Yongli Yan, Ze Xiang Shen, Kian Ping Loh*,and Ding Yuan Tang**

[*] Prof. K. P. Loh. Corresponding-Author, Dr. Q. Bao, Dr. Y. Wang, Dr. Y. Yan
Department of Chemistry, National University of Singapore,
3 Science Drive 3, Singapore 117543 (Singapore)
E-mail: chmlohkp@nus.edu.sg

Prof. D. Y. Tang, H. Zhang
School of Electrical and Electronic Engineering, Nanyang Technological University,
Singapore 639798 (Singapore)
E-mail: edytang@ntu.edu.sg

Prof. Z. X. Shen, Dr. Z. Ni
School of Physical and Mathematical Sciences, Nanyang Technological University
Singapore 639798 (Singapore)

[**] This project is funded by NRF-CRP award "Graphene and Related Materials and Devices". Supporting Information is available online from Wiley InterScience or from the author.

**Abstract**

The optical conductance of monolayer graphene is defined solely by the fine structure constant, $\alpha = e^2/\hbar c$ (where e is the electron charge, ħ is Dirac's constant and c is the speed of light). The absorbance has been predicted to be independent of frequency. In principle, the interband optical absorption in zero-gap graphene could be saturated readily under strong excitation due to Pauli blocking. Here, we demonstrate the use of atomic layer graphene as saturable absorber in a mode-locked fiber laser for the generation of ultrashort soliton pulses (756 fs) at the telecommunication band. The modulation depth can be tuned in a wide range from 66.5% to 6.2% by varying the thickness of graphene. Our results suggest that ultrathin graphene films are potentially useful as optical elements in fiber lasers. Graphene as a laser mode locker can have many merits such as lower saturation intensity, ultrafast recovery time, tunable modulation depth and wideband tuneability.

## 1. Introduction

The electronic and optical properties of graphene, a single atomic layer of $sp^2$-hybridized carbon forming a honeycomb crystal lattice, can be described in terms of massless Dirac fermions with linear dispersion near the Fermi energy. The optical interband transitions are expected to be frequency independent and solely determined by the optical conductance in a broad range of photon energies. The remarkably large absorption of atomic layer graphene implies lower saturation intensity or higher photocarrier density compared to traditional semiconductor materials such as gallium arsenide (GaAs). For example, a single graphene sheet was calculated to absorb a significant fraction ($\pi\alpha$ = 2.3 %) of incident infrared-to-visible light,[1-5] compared to about 1% absorption of near-gap light for 10 nm thick GaAs quantum well.[6] This means that in principle, graphene can be saturated readily under strong excitation from visible to near-infrared region due to the universal optical absorption and zero band gap. This has relevance for the mode locking of fiber lasers, where wideband tuneability may be obtained using graphene as the saturable absorber.

Figure 1a shows excitation processes responsible for absorption of light in monolayer graphene, in which electrons from the valence band (orange) are excited into the conduction band (yellow). Shortly after photo-excitation (within 10-150 fs [7]), these hot electrons thermalize and cool down to form a hot Fermi-Dirac distribution (Figure 1b) with electronic temperature $T_e$.[8, 9] These newly created electron-hole pairs could block some of the originally possible interband optical transitions in a range of $k_BT_e$ ($k_B$ is the Boltzmann constant) around the Fermi energy $E_F$ and decrease the absorption of photons $\hbar\omega \sim k_BT_e$. In the following ~ 1 picosecond, intraband phonon scattering further cools the thermalized carriers. After that, electron-hole recombination will dominate the process until the equilibrium electron and hole distribution is restored (Figure 1b).[7, 10] However, these only describe the linear optical transition under low excitation intensity. As the excitation is increased to higher intensity, the photogenerated carriers increase in concentration (much larger than the intrinsic electron and hole carrier densities of ~ $8 \times 10^{10}$ $cm^{-2}$ in graphene at room temperature[10]) and cause the states near the edge of the conduction and valence bands to fill, blocking further absorption (Figure 1c), and thus imparting transparency to light at photon energies just above the band-edge. Band-filling occurs because no two electrons can fill the same state. Thus, saturable absorption or absorption bleaching is achieved due to this *Pauli blocking* process.[11]

In this paper, we investigate the nonlinear optical properties of large area atomic layer graphene thin films transferred onto optical fiber. We show that graphene has relatively lower saturation adsorption intensity but much larger modulation depth compared to traditional saturable absorbers. For the first time, we demonstrate the use of atomic layer graphene as saturable absorber in a mode-locked fiber laser for the generation of ultrashort soliton pulses at the telecommunication band.

## 2. Results and Discussion

### 2.1. Characterization of Graphene

Chemical vapor deposition (CVD) was used[12, 13] to synthesize large area graphene thin films (see Experimental and Supporting Information, Section 1 and 2). The CVD deposited graphene is highly crystalline, continuous, and could be readily lifted off and transferred to an optical fiber pigtail end (Figure 2a, b). The thickness of the graphene film could be controlled by the growth time. Electron diffraction pattern (Section 3 in Supporting Information) of the CVD graphene reveals that its stacking geometry is different from the hexagonal AB (Bernal) stacking in graphite, resulting in an effective decoupling of adjacent layers because of misoriented stacking. Theoretical simulations have demonstrated that[14-18] such stacking disorder helps to preserve the electronic structure of single layer graphene in multilayers.

Raman spectroscopy was combined with optical contrast spectroscopy[19, 20] to identify the thickness of graphene layers covered on the fiber core area (Figure 2c-g). (see Supporting Information, Section 4). This allows us to unambiguously relate the saturable absorption strength of the graphene with the thickness of the graphene film transferred onto the optical fiber. Figure 2c shows the Raman spectra of graphene with different number of layers. The prominent G band at 1584 $cm^{-1}$ and 2D band at 2689 $cm^{-1}$ are clearly resolved. The relatively weak D band at 1344 $cm^{-1}$ indicates a low density of defects and high crystallinity of the CVD grown graphene. As discussed, the misoriented layers in our graphene are different from AB stacked graphite and each layer maintains the integrity of single layer graphene. For example, a five-layered graphene has narrow and symmetric 2D band with width of 28.8 $cm^{-1}$ (inset of Figure 2c), comparable to monolayer graphene.[21] We even observed G band resonance on some of the graphene films whose signal is tens of times stronger than those without resonance. It means that the Raman intensity from these graphene films are not directly and simply correlated to the thickness of graphene films. Therefore, we use optical contrast to identify the graphene film thickness as the elastic scattering is not or less affected by the electronic structure of graphene or few layers graphene.

As the basic component materials of fiber core and quartz are very similar ($SiO_2$), we use the Raman and contrast spectra from graphene on quartz substrate as references. The contrast between graphene layers and the quartz substrate, which makes graphene visible, was generated from a reflection spectrum using a normal white light source. The contrast spectra $C(\lambda)$ are defined using the following equation,

$$C(\lambda) = \frac{R_0(\lambda) - R(\lambda)}{R_0(\lambda)} \qquad (1)$$

where $R_0(\lambda)$ is the reflection spectrum from substrate, and $R(\lambda)$ is the reflection spectrum from graphene sheets.[19, 20] The contrast values of graphene sheets on quartz substrate are found to be around M × (-0.068) for Mth (M = 1, 2, 3,…) layers graphene on quartz substrate. Figure 2d shows the contrast spectra of graphene with different number of graphene layers. With this technique, the thickness of an unknown graphene sheet can be determined directly by comparing the contrast value with these standard values. Figure 2g shows a representative optical contrast image with 2~4 layers graphene coated on the core of a fiber. Due to inhomogeneous coverage within the fibre core, the number of layers are specified within +/-1.

### 2.2. Nonlinear Optical Properties

The optical nonlinearities are directly related to photocarrier density. A simple two-level saturable absorber model[22, 23] widely used for two dimensional quantum wells was adapted to describe the nonlinear saturable absorption in graphene with approximate form

$$\alpha^*(N) = \frac{\alpha^*_S}{1 + N/N_S} + \alpha^*_{NS} \qquad (2)$$

where $\alpha^*(N)$ is the absorption coefficient, and $\alpha^*_S$ and $\alpha^*_{NS}$ is the saturable and nonsaturable absorption components. $N$ is the photoinduced electron-hole density, and $N_s$, the saturation density, is the value of $N$ for which the absorption falls to one-half of its initial value. The photocarrier density described by the rate equation (see Equation (7) in Section 5 of Supporting Information) can be simplified by introducing intensity ($I$) of continuous wave (CW) or long-pulse excitation

$$N = \frac{\alpha^* I \tau}{\hbar\omega} \qquad (3)$$

where τ is the carrier recombination time and ω is the frequency of light.Equation (3) reveals a speed-intensity tradeoff, i.e., the longer the time before recombination, the less CW intensity is required to achieve the same carrier density.[22, 23]

Figure 3a shows the total absorption of graphene films as a function of the number of layer using phonon energies 0.75 ~ 0.85 eV. The result agrees with the Beer-Lambert law with a constant absorbance of $\pi\alpha$ for monolayer graphene sheet.[1-5] The nonlinear saturable absorption properties of multilayers graphene are studied by power-dependent measurements at 1550 nm, as shown in Figure 3b. The saturation absorption could be seen quite clearly with the increase of incident light intensity. If $N$ in equation (2) is substituted by equation (3), the absorption as a function of increasing light intensity can be fitted by

$$\alpha^*(I) = \frac{\alpha^*_s}{1 + I/I_S} + \alpha^*_{NS} \qquad (4)$$

where $I_s$ is the saturation intensity, defined as the optical intensity required in a steady state to reduce the absorption to half of its unbleached value. This yields a saturation intensity ranging from 0.71 to 0.61 MW $cm^{-2}$ (See Supporting Information, Section 6) when the number of graphene layer is varied from 3±1 to 10±1. Meanwhile, the modulation depth is reduced from 66.5% to 6.2% (Figure 3c) due to the increased nonsaturable loss caused by enhanced scattering of graphene multi-layers. Compared to single-walled carbon nanotubes (SWNTs)[24-26] and semiconductor saturable absorber mirrors (SESAMs),[27] the saturation intensity of graphene is one order of magnitude lower while the modulation depth is 2~3 times larger. The relatively large modulation depth of the atomic layer graphene results from its much smaller nonsaturable loss, which is an intrinsic merit of graphene arising from its two-dimensional (2D) structure. In the case of SWNTs, the presence of metal catalyst and scattering from bundled CNT contributes significantly to the nonsaturable loss.

As the ultrafast carrier kinetics is not very sensitive to the number of graphene layers due to the loosely bonded layers of graphene-like planes,[28] for simplicity we assume that graphene films of different thickness have similar carrier recombination time,[7, 10] $\tau$ = 1.67 ps ($\tau_2$ in Figure 3d). The saturated carrier density $N_s$ in the different graphene layers can then be estimated using equation (2). For 2 ~ 4 layers graphene, $I_s$ = 0.71 MW $cm^{-2}$, it gives $N_s$ =5.84 $\times 10^{13}$ $cm^{-2}$; for 9~11 layers graphene, $I_s$ = 0.61 MW $cm^{-2}$, it gives $N_s$ =8.16 $\times 10^{14}$ $cm^{-2}$, which is more than 4 orders of magnitude higher than the intrinsic carrier density in graphene. Figure 3c plots the saturated carrier density versus number of graphene layers. The saturated carrier density of 9 ~ 11 layers graphene is about 3 orders of magnitude higher than that of traditional SESAMs with comparable modulation depth. For example, for 10 nm thick GaAs quantum well, it only gives $N_s$ =2.4 × $10^{11}$ $cm^{-2}$ (See Supporting Information, Section 5).

### 2.3. Graphene Mode Locked Fiber Laser

Figure 4 shows the experimental setup of the graphene mode locked fiber laser. The basic operation of the laser is described in the experimental section. Neither intra-cavity real polarizer nor similar polarization dependent components was used in the cavity to exclude the mode locking states induced by nonlinear polarization rotation (NPR).It is well known that mode-locked lasers emit stable and ultrashort pulses if the soliton pulse shaping caused by the balance between the self-phase modulation (SPM) and the negative group velocity dispersion (GVD) is incorporated in the laser action. Interestingly, we found that the atomic layer graphene exhibited extremely large normal dispersion in comparison with SESAMs or SWNTs (See Supporting Information, Section 7). The extreme large normal dispersion of the atomic layer graphene might correlate with the strong electron-photon interaction in the two-dimensional lattice.[29] Thus, in order to obtain unambiguous evidence of soliton mode locking of our fiber laser, i.e., the appearance of solitonic sibebands, 100 m of extra SMF was added to compensate for the normal dispersion of graphene so that the net cavity dispersion becomes anomalous.[23]

Figure 5 summarizes the characteristics of the soliton pulses emitted from the fiber laser mode locked with the atomic layer graphene film. The pulse train of the laser output depicted in Figure 5a has a repetition rate of 1.79 MHz, which matches with the cavity roundtrip time and verifies that the laser is mode locked. Figure 5b shows the optical spectrum of the mode locked pulses. Its central wavelength is at 1565 nm and it has a bandwidth of 5 nm. The solitonic sidebands are clearly visible on the spectrum. The formation of solitonic sidebands is a result of periodic soliton perturbation in the laser cavity, which is a characteristic of the soliton operation of the fiber lasers.[23, 27] We also measured the radio-frequency (RF) spectrum of the mode-locked pulses. Its fundamental peak locates at the cavity repetition rate (1.79 MHz), as shown in Figure 5c, with a signal-to-noise ratio of 65 dB ($10^6$ contrast). It indicates the stability of the soliton pulses. The autocorrelation (AC) trace of the mode-locked pulses is shown in Figure 5d, which is well fitted by a $\text{sech}^2$ profile with 1.17 ps full-width at half-maximum (FWHM). The real pulse width is then obtained by multiplying the AC trace width with the de-correlation factor, which is 0.648 for the $\text{sech}^2$ pulses. This gives a pulse width of 756 fs. An output power up to 2 mW was recorded with a slope efficiency up to 3% (Figure 5e). The result is comparable to the SWNTs mode locked fiber lasers [25, 30] but it could be further improved by laser cavity design. The pump threshold for self-starting of the mode locking ranges from 8 ~ 40 mW.

### 2.4. Simulation of Laser Operation

In order to confirm the measured saturable absorption properties of the atomic layer graphene, we further numerically simulated the laser operation using the actual fiber parameters and the measured graphene parameters. Our simulation was based on a laser roundtrip model that takes account of the pulse propagation in the optical fibers, which is described by coupled Ginzburg-Landau equations[31] (Supporting Information, Section 7), and the actions of the individual cavity components. In our simulations the large local dispersion of the graphene film was modeled by multiplying a dispersion matrix with a normal dispersion value that equals to that of 87 meters SMF. Through slightly adjusting the cavity parameters including the cavity birefringence and the pumping strength, stable mode locking could indeed be obtained. Figure 5f shows an example of the calculated soliton state, obtained with gain = 28.9 dB and a very weak cavity birefringence of $L/L_b = 0.2$. The optical spectrum and pulse profile of the numerically calculated solitons are shown in the insets of Figure 5f. The separation between the first Kelly sidebands and the soliton peak is about 5.78 nm, the 3 dB spectral bandwidth of the pulses is about 5.4 nm, and the FWHM pulse width is about 1.16 ps, which are quantitatively consistent with our experimental observations.

## 3. Conclusions

In summary, we discovered that the atomic layer graphene film could exhibit saturable absorption. Moreover, its saturable absorption could be achieved at a much lower saturation threshold compared to materials like carbon nanotubes. The saturation absorption strength could be modulated over a wide range by varying the number of graphene layers. The unique properties of graphene with low nonsaturable loss render graphene a promising ultrathin saturable absorber for ultrafast fiber lasers. New generation low-noise and inexpensive light sources can be developed with graphene for applications in optical communications.

## 4. Experimental

*Large area graphene*: The large area graphene films was produced by chemical vapor deposition (CVD) method.[12, 13] In a typical experiment, a $SiO_2$/Si substrate with 300 nm Ni film was loaded into a CVD chamber. The Ni catalyst was activated at 700 °C in 100 sccm $H_2$ gas flow. The samples were heated up to 900 ~ 1000 °C inside a quartz tube under the flow of Ar/$CH_4$/$H_2$ mixture flow (Ar: $CH_4$: $H_2$: = 3: 1: 1) and reacted for 10 min. Finally, the system was allowed to rapidly cool down to room temperature at the rate of ~ 10 °C /s under the protection of Ar gas flow. An ultrathin graphene layer was deposited on the Ni surface upon cooling of the sample.

*Optical Measurements:* A weak CW from a distributed feedback (DFB) laser with linewidth narrower than 0.01 nm was used as the input light source to investigate the linear absorption characteristics of graphene with different layers. A stable and standard soliton mode locked fiber laser with output pulse width 1 ps and repetition rate 100 MHz was used as input seed pulse for nonlinear absorption measurements (see Supporting Information, Section 6). The seed pulses were amplified through an erbium doped fiber amplifier (EDFA) followed by a coupler to split the light into two patchcords for sample testing. To reduce the errors caused by the fluctuation of the input light source, the input power (not passing through graphene) and output power (passing through graphene) were determined simultaneously through two set of power meters with equal optoelectronic responses and parameter setups including sweeping time. The time-resolved pump-probe profiles of graphene films were obtained using a femtosecond Ti:sapphire amplifier (Spectra Physics, Spitfire) pumped by an Empower (Spectra Physics) with pump laser operating at 400 nm, probe laser at 630 nm, and with a pump power of 30 μW. Pulse duration in pump-probe experiments was 50 fs and instrumental response time was less than 80 fs.

*Ultrafast Laser Experimental Setup:* The fiber laser has a ring cavity which consists of a piece of 6.4 m EDF with group velocity dispersion (GVD) of 10 ps/km/nm, 105.3 m SMF with GVD 18 ps/km/nm. Solitonic sidebands could be observed on the spectra of the mode locked pulses, demonstrating that the net cavity dispersion is anomalous in the present cavity. The total fiber dispersion is about 1.96 ps/nm. A 10% fiber coupler was used to output the signal, and the laser was pumped by a high power fiber Raman laser source (BWC-FL-1480-1) of wavelength 1480 nm, which has been frequently used in the previous soliton fiber laser works.[31] A polarization independent isolator was spliced in the cavity to force the unidirectional operation of the ring. An intra cavity polarization controller was used to change the cavity linear birefringence. Graphene mode-locker was placed between the output coupler and polarization controller.

**Figure Caption**

**Figure 1.** Absorption of light in graphene. a) Schematic excitation process responsible for absorption of light in graphene. The arrow indicates optical interband transition. b) The photogenerated carriers thermalize and cool down within subpicosecond to form a hot Fermi-Dirac distribution, an equilibrium electron and hole distribution could be finally approached through intraband phonon scattering and electron-hole recombination. c) At enough high excitation intensity, the photogenerated carriers cause the states near the edge of the conduction and valence bands to fill, blocking further absorption.

**Figure 2.** a) Photograph of graphene floating on de-ionized (DI) water. The red arrows show two cut graphene flakes and the yellow dash arrows indicate the trace after graphene flakes float off from the $Ni/SiO_2$ substrate. b) Photograph of fiber pigtail coated with graphene film and schematic of the end of optical fiber pigtail with a graphene film coating on pinhole (yellow). c**)** Raman spectra of graphene sheets with different thickness. Inset shows

Lorentzian fitting of 2D band of 5 layers graphene. d) Contrast spectra of graphene sheets with different thickness. The contrast values were counted in the spectrum range of 550 to 650 nm. e) Optical image of fiber pinghole (diameter: ~ 125 μm) coated with large area graphene. The scale bar is 20 μm. f) Raman image around fiber core plotted by the intensity of G band. The scale bar is 1 μm. g) Optical contrast image showing 2~4 layers graphene coated on fiber core. The scale bar is 1 μm.

**Figure 3.** a) Linear absorption of graphene films with different number of layers (excited at phonon energies 0.75~0.85 eV). b) Nonlinear absorption of graphene films with different number of layers. The dots are the experimental data and the solid curves are analytical fits to the data using equation (4). c) Modulation depth and saturated carrier density versus number of graphene layers. d) Measured transmittivity transients for multilayer graphene. The open circles are the experimental data and the solid curve is analytical fit to the data using exponentials with time constants $\tau_1$ and $\tau_2$. The fast relaxation time $\tau_1$ corresponds to carrier-carrier intraband scattering rates and the slow relaxation time $\tau_2$ correlates with electron-hole interband recombination.

**Figure 4.** Laser configuration constituting a ring cavity. This schematic shows the standard fiber-optic components such as wavelength division multiplexer (WDM), polarization controller, coupler, optical isolator, erbium doped fiber (EDF) and single mode fiber (SMF). The detail parameters and the basic laser operation are discussed in the Methods.

**Figure 5.** Mode-locking characteristics. a) Typical laser output pulse train. b) Output pulse spectrum, centered at 1567 nm, with solitonic sidebands. c) Fundamental of a typical RF spectrum of the laser output after optical-to-electrical conversion. d) AC trace of laser output

and $sech^2$ fitting curve. e) Output power as a function of pump power. f) Numerically calculated pulse evolution with cavity round trips. The insets show output pulse spectrum (upper left) and profile (upper right) of the numerically calculated solitons.

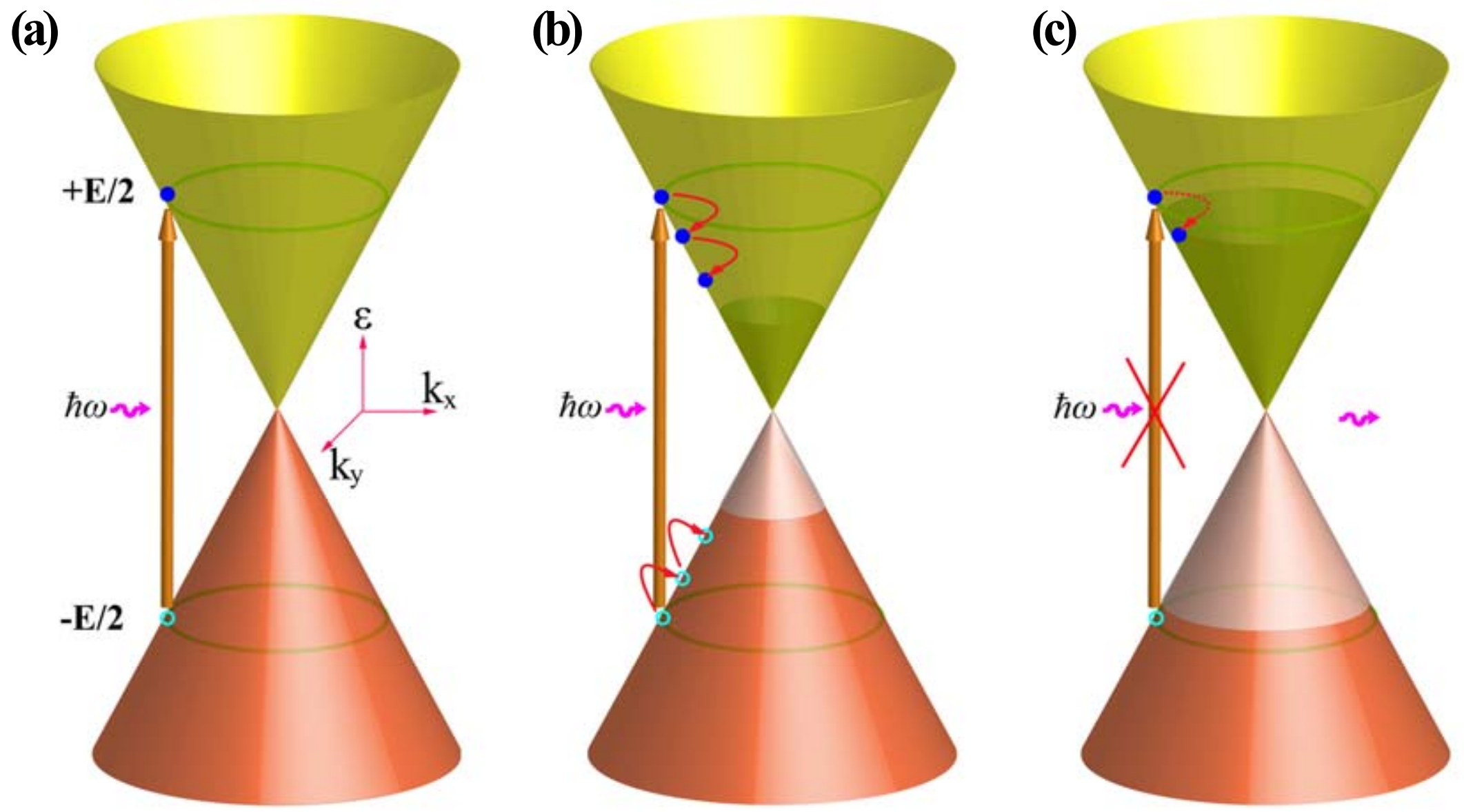

**Figure 1.** Absorption of light in graphene. a) Schematic excitation process responsible for absorption of light in graphene. The arrow indicates optical interband transition. b) The photogenerated carriers thermalize and cool down within subpicosecond to form a hot Fermi-Dirac distribution, an equilibrium electron and hole distribution could be finally approached through intraband phonon scattering and electron-hole recombination. c) At enough high excitation intensity, the photogenerated carriers cause the states near the edge of the conduction and valence bands to fill, blocking further absorption.

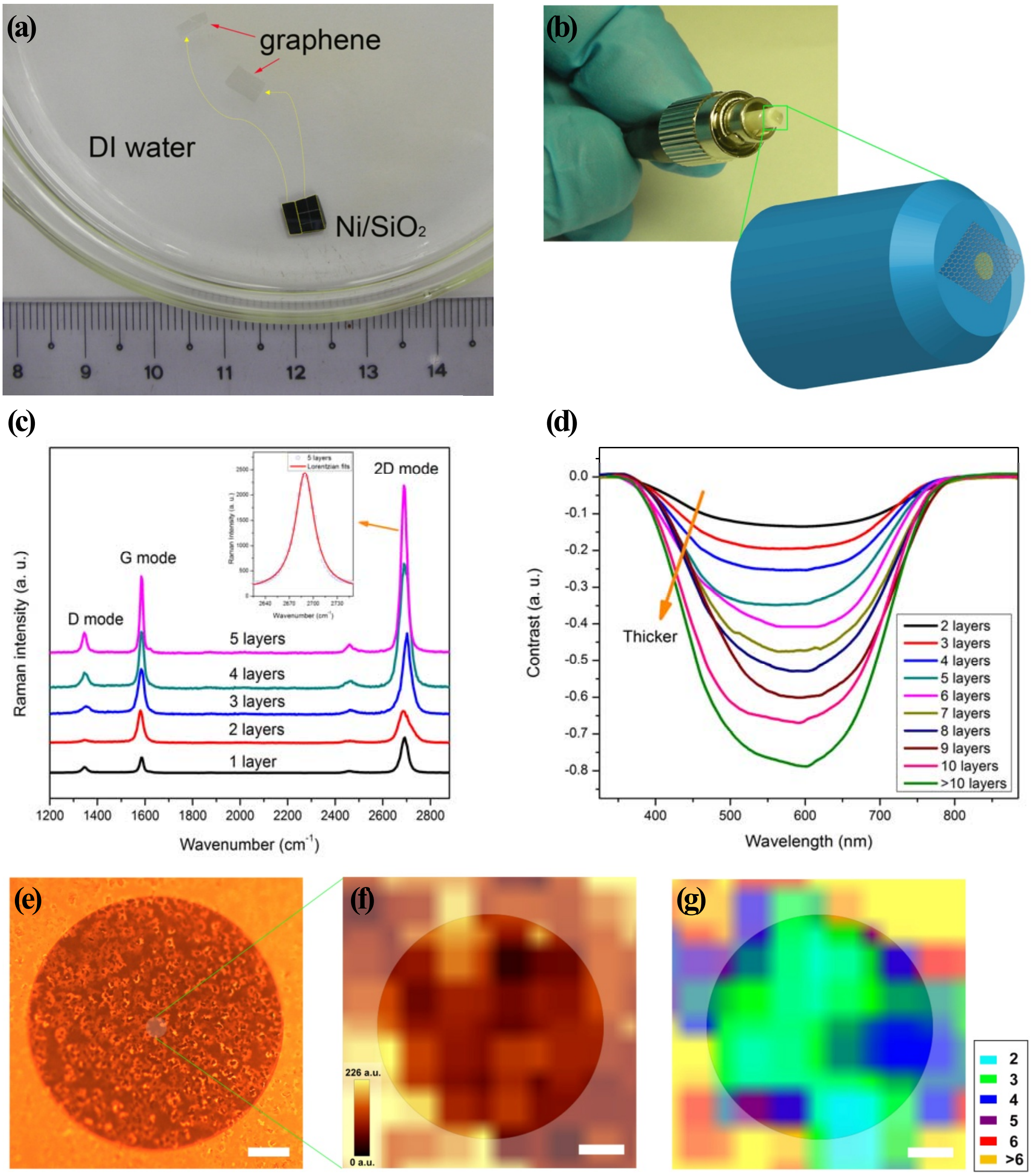

**Figure 2.** a) Photograph of graphene floating on de-ionized (DI) water. The red arrows show two cut graphene flakes and the yellow dash arrows indicate the trace after graphene flakes float off from the $Ni/SiO_2$ substrate. b) Photograph of fiber pigtail coated with graphene film and schematic of the end of optical fiber pigtail with a graphene film coating on pinhole (yellow). c) Raman spectra of graphene sheets with different thickness measured on quartz. Inset shows Lorentzian fitting of 2D band of 5 layers graphene. d) Contrast spectra of graphene sheets with different thickness measured on quartz. The contrast values were counted in the spectrum range of 550 to 650 nm. e) Optical image of fiber pinghole (diameter: ~ 125 μm) coated with large area graphene. The scale bar is 20 μm. f) Raman image around fiber core plotted by the intensity of G band. The scale bar is 1 μm. g) Optical contrast image showing 2~4 layers graphene coated on fiber core. The scale bar is 1 μm.

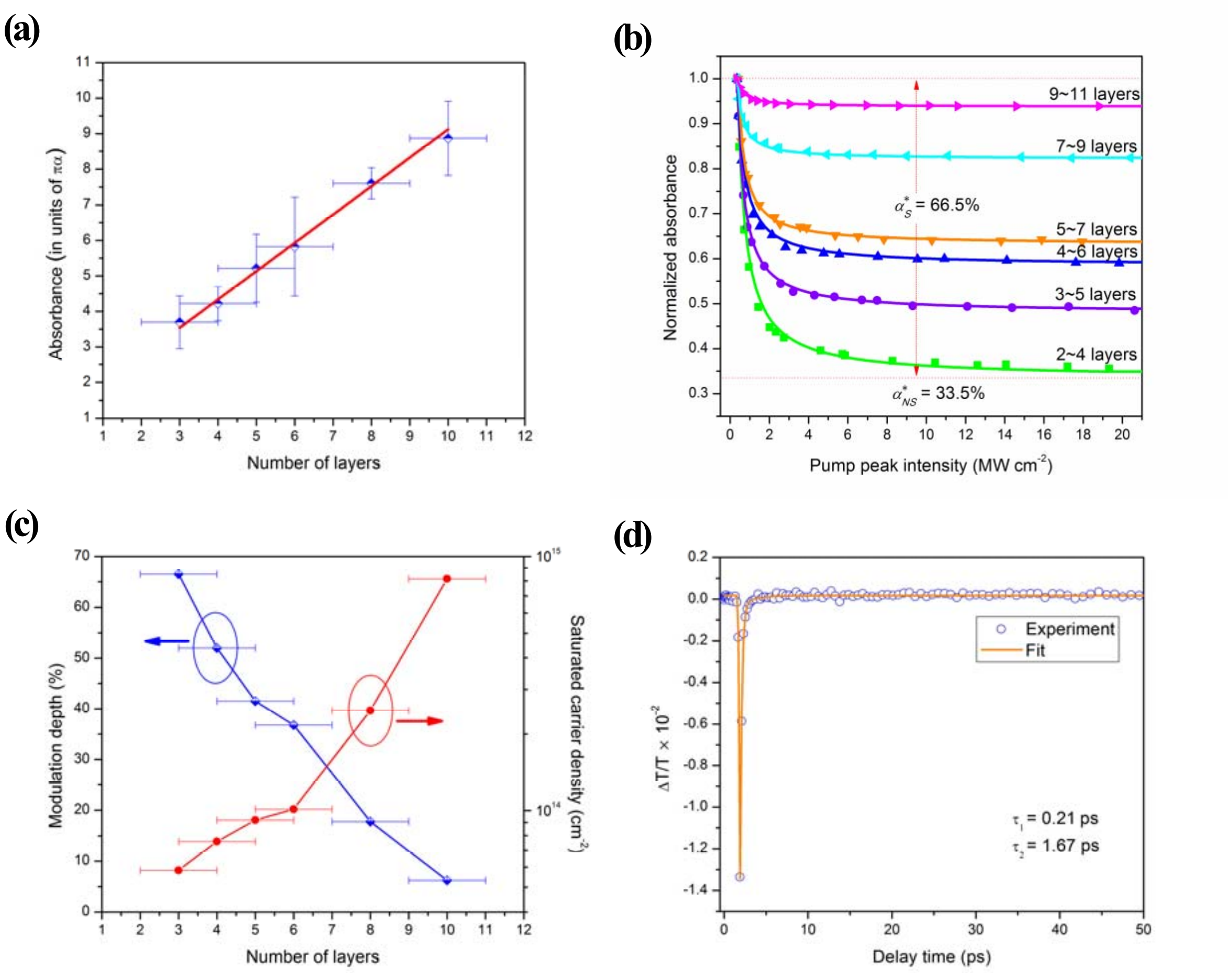

**Figure 3.** a) Total absorption of graphene films with different number of layers (excited at phonon energies 0.75~0.85 eV). b) Nonlinear absorption of graphene films with different number of layers. The dots are the experimental data and the solid curves are analytical fits to the data using equation (4). c) Modulation depth and saturated carrier density versus number of graphene layers. d) Measured transmittivity transients for multilayer graphene. The open circles are the experimental data and the solid curve is analytical fit to the data using exponentials with time constants $\tau_1$ and $\tau_2$. The fast relaxation time $\tau_1$ corresponds to carrier-carrier intraband scattering rates and the slow relaxation time $\tau_2$ correlates with electron-hole interband recombination.

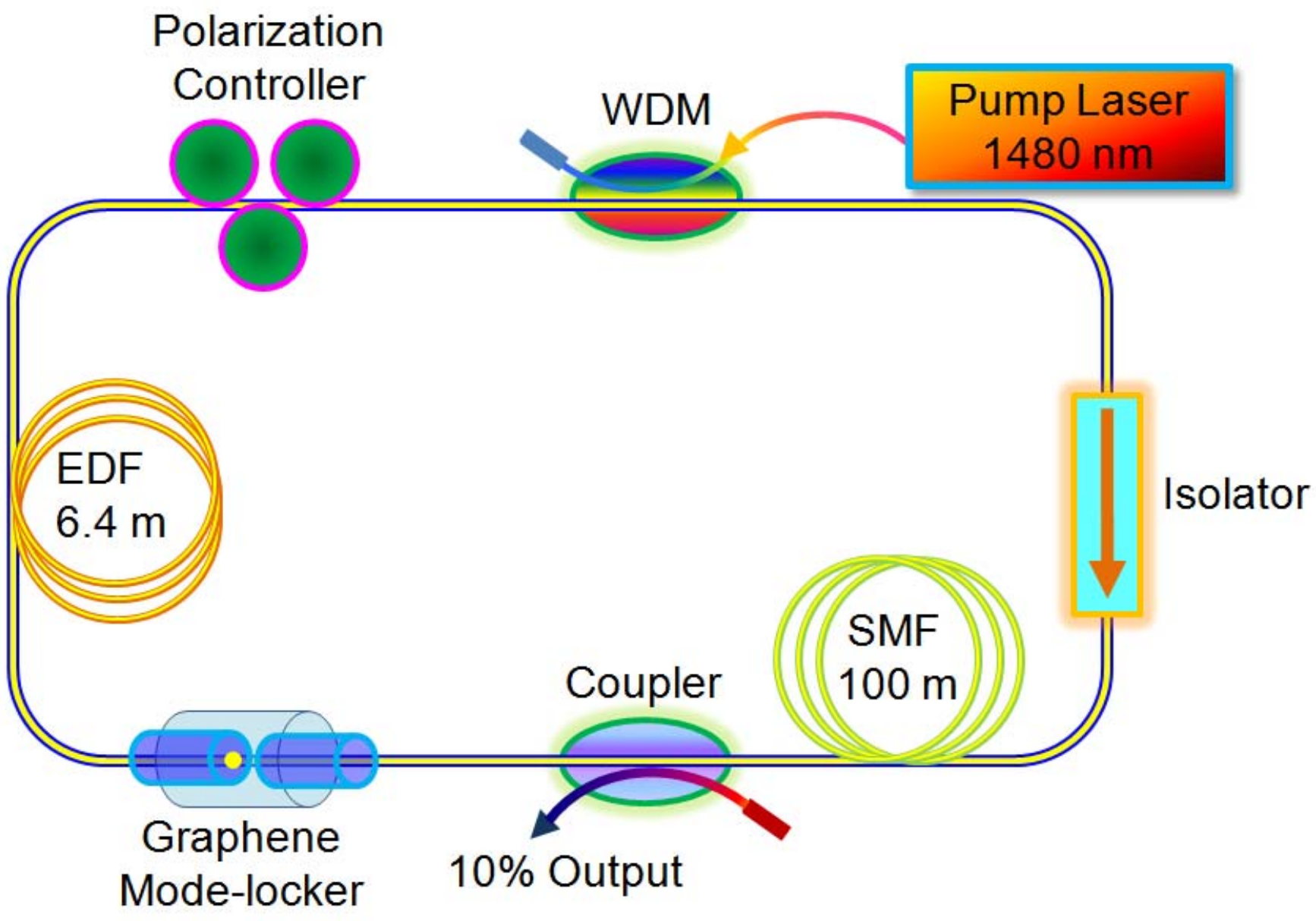

**Figure 4.** Laser configuration constituting a ring cavity. This schematic shows the standard fiber-optic components such as wavelength division multiplexer (WDM), polarization controller, coupler, optical isolator, erbium doped fiber (EDF) and single mode fiber (SMF). The detail parameters and the basic laser operation are discussed in the Methods.

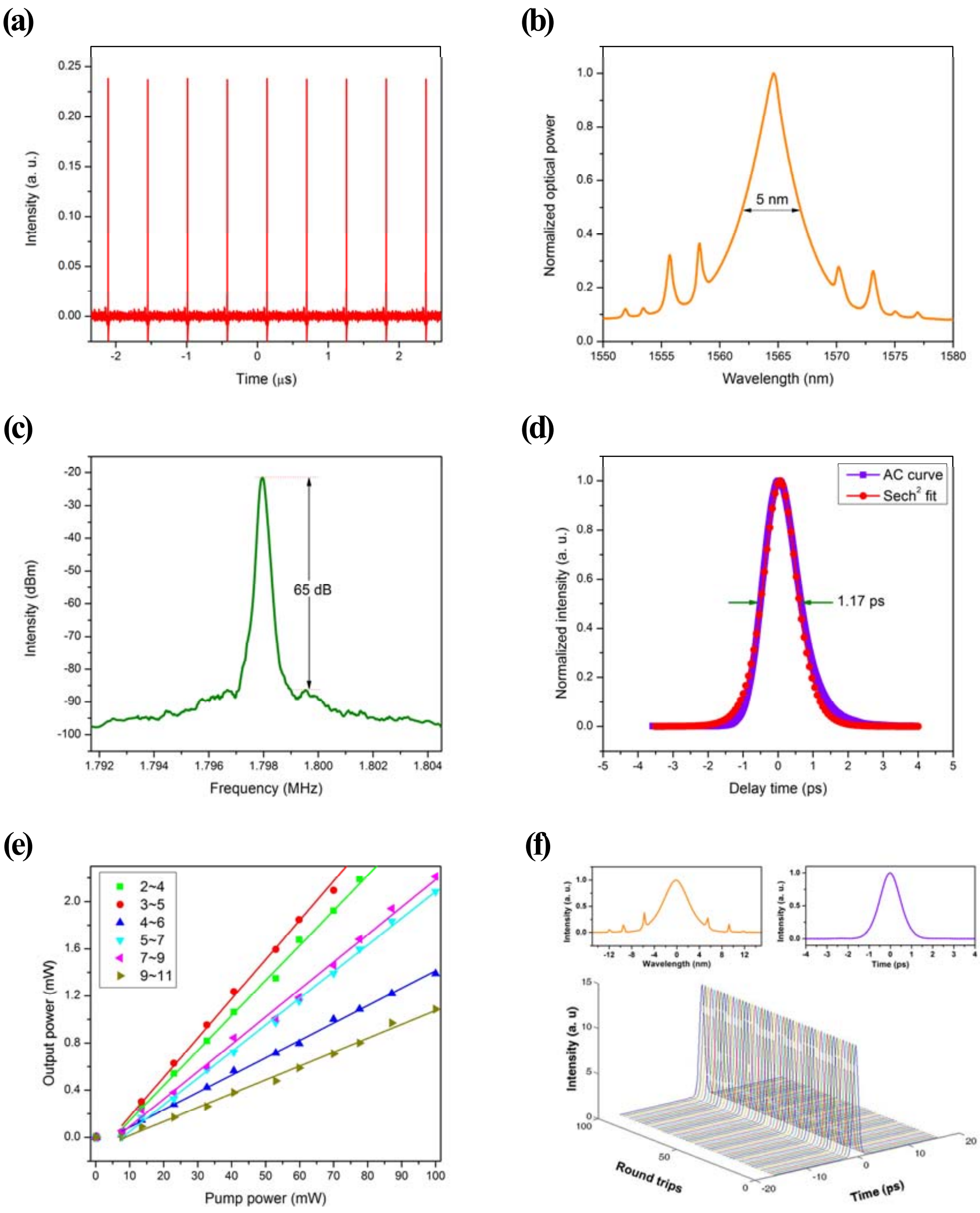

**Figure 5.** Mode-locking characteristics. a) Typical laser output pulse train. b) Output pulse spectrum, centered at 1567 nm, with solitonic sidebands. c) Fundamental of a typical RF spectrum of the laser output after optical-to-electrical conversion. d) AC trace of laser output and sech$^2$ fitting curve. e) Output power as a function of pump power. f) Numerically calculated pulse evolution with cavity round trips. The insets show output pulse spectrum (upper left) and profile (upper right) of the numerically calculated solitons.

**The table of contents entry**

we demonstrate the use of atomic layer graphene as saturable absorber in a mode-locked fiber laser for the generation of ultrashort soliton pulses at telecommunication band. Our results suggest that ultrathin graphene films are potentially useful as optical elements in fiber lasers.

Authors: Qiaoliang Bao, Han Zhang, Yu Wang, Zhenhua Ni, Yongli Yan, Ze Xiang Shen, Kian Ping Loh*,and Ding Yuan Tang*

Title: Atomic layer graphene as saturable absorber for ultrafast pulsed lasers

ToC figure ((55 mm broad, 50 mm high, or 110 mm broad, 20 mm high))

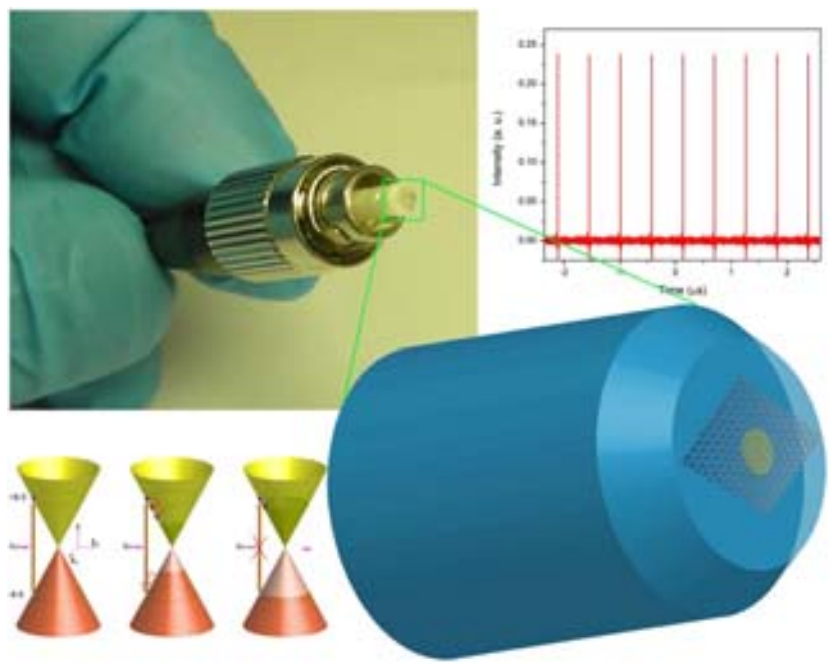

Column Title: *Qiaoliang Bao et al.* Graphene based laser